\documentclass[aps,twocolumn,groupedaddress,nofootinbib]{revtex4}
\usepackage{graphicx}
\usepackage{amssymb,amsmath}
\usepackage[usenames,dvipsnames]{color}
\usepackage{slashed}

\def\Re{\mathop{\rm Re}\nolimits}
\def\Im{\mathop{\rm Im}\nolimits}
\newcommand{\hc}{{\rm h.c.}}
\newcommand{\ft}[2]{{\textstyle\frac{#1}{#2}}}
\def\rmi{{\rm i}}

\newcommand{\PLa}{} % to reinstall, change the last to {P_L}
\newcommand{\PRa}{} % to reinstall, change the last to {P_R}
\newcommand{\gamfive}{\gamma_5}
\newcommand{\be}{\begin{equation}}
\newcommand{\ee}{\end{equation}}
\newcommand{\bea}{\begin{eqnarray}}
\newcommand{\eea}{\end{eqnarray}}	

\newcommand{\ba}{\begin{array}}
\newcommand{\ea}{\end{array}}

\newcommand{\mgr}{m_{3/2}}

\definecolor{MyDarkGray}{RGB}{140,140,140}

\begin{document}
\begin{flushleft}
KCL-PH-TH/2013-01 \\
LCTS/2012-34 \\
%CERN-PH-TH/2012-xxx\\
\end{flushleft}

\title{On Gravitino properties  in a Conformal Supergravity Model}

\author{Nick E. Mavromatos}
\affiliation{Theoretical Particle Physics and Cosmology Group, Department of Physics, King's College London, Strand, London WC2R 2LS, UK; \\
Theory Division, Physics Department, CERN, CH-1211  Geneva 23, Switzerland}

\author{Vassilis C. Spanos} 
\affiliation{Institute of Nuclear Physics, NCSR  ``Demokritos'', GR-15310 Athens, Greece}

\begin{abstract} 
In the context of a conformal Supergravity (SUGRA) model in the Einstein frame, in which the (next to) minimal supersymmetric standard model can embedded naturally to produce chaotic inflation scenarios, we 
study properties of gravitino in the cases where it is stable or unstable. 
In the latter case, we demonstrate that for large dilaton scale factors there is an enhanced magnitude of the gravitino width, when it decays to neutralino dark matter, as compared with the standard SUGRA case. In this context, we discuss the associated consequences as far as Big Bang Nucleosynthesis constraints and avoidance of gravitino overproduction are concerned. 
\end{abstract}

\maketitle

\section{Introduction} 

In \cite{Ferrara:2010yw,Ferrara:2010in} simple classes of supergravity (SUGRA) models 
describing superconformal coupling of matter to supergravity have been considered. The models contain non-minimal 
scalar/space-time-curvature couplings of the form $\Phi \, R$, where $\Phi$ is a frame function, depending in general on matter supermultiplets, including dilatons. Such couplings have been argued to lead naturally to 
Higgs-inflation in both non-supersymmetric~\cite{Bezrukov:2007ep,Bezrukov:2009db,Bezrukov:2010jz,GarciaBellido:2011de}
and supersymmetric theories~\cite{Einhorn:2009bh,AlvarezGaume:2010rt,AlvarezGaume:2011xv}.
Scale-free globally supersymmetric theories, such as the Next to Minimal Supersymmetric Standard Model (NMSSM)~\cite{NMSSM} with a scale-invariant
superpotential, can be naturally embedded~\cite{Ferrara:2010in,Einhorn:2009bh} into this class of theories, leading to new classes of chaotic inflationary scenarios~\cite{Ferrara:2010yw}.  Moreover, such models have been considered in \cite{Ellis:2011mz} in connection with the possibility of dynamical breaking of supergravity theories, exploring further the conformal couplings of the gravitino four-fermion interactions. 

It is the point of this article to discuss the properties of gravitino fields in such models, in particular in the context of the NMSSM. Specifically, we shall analyse  decay processes involving gravitinos and calculate the corresponding life time. Depending on the strength of the conformal couplings, the width can be suppressed or enhanced significantly. In the case of enhancement the rapid decay of the gravitino implies a resolution of the gravitino overproduction, avoiding the Big-Bang-Nucleosynthesis (BBN) constraints. 

The structure of the article is as follows: in the next section \ref{sec:2}, we describe the basic Lagrangian formalism and properties underlying the conformal supergravity models of  
\cite{Ferrara:2010yw,Ferrara:2010in}. In section \ref{sec:3} we analyse the main decay processes involving gravitinos, calculate the associated widths (and lifetimes), and discuss how the latter are constrained by BBN. Conclusions and outlook are presented in section \ref{sec:4}.

\section{Lagrangian Formalism of Conformal Supergravity Models \label{sec:2}}

The action of the conformal supergravity models of  \cite{Ferrara:2010yw,Ferrara:2010in}, that we shall employ in our analysis below, in the Jordan frame reads:
\begin{eqnarray}
&&  e^{-1}\mathcal{L} =-\ft16\Phi \left[ R(e)-\bar \psi_\mu R^\mu\right] -\ft16
(\partial _\mu\Phi )(\bar \psi \cdot \gamma \psi ^\mu) +\nonumber\\
&&+\mathcal{L}_0+\mathcal{L}_{1/2}+\mathcal{L}_1-V+\mathcal{L}_m+\mathcal{L}_{\mathrm{mix}}+\mathcal{L}_d+\mathcal{L}_{4\mathrm{f}}\,,
 \label{Poincact}
\end{eqnarray}
where the curvature $R(e)$ uses the torsionless connection $\omega _\mu {}^{ab}(e)$, with $e_a^\mu$ the vielbeins, and $e$ the vielbein determinant, 
and the gravitino kinetic term is defined using
\begin{equation}
R^\mu \equiv \gamma ^{\mu \rho \sigma }\left( \partial _{\rho }+\ft14\omega _{\rho }{}^{ab}(e)\gamma _{ab} -\ft32
\rmi\mathcal{A}_\rho \gamfive\right)\psi _\sigma \,.
 \label{defRmuPoinc}
\end{equation}
Here $\mathcal{A}_\mu $ is the part of the auxiliary vector field
containing only bosons, namely:
\begin{equation}
   \mathcal{A}_\mu =\ft16\rmi\left(
\partial _\mu z^\alpha \partial _\alpha \mathcal{K}
-\partial _\mu \bar z^{\bar \alpha }\partial _{\bar \alpha }\mathcal{K}\right)
-\ft13 A_\mu {}^AP_A\,,  \label{Amu}
\end{equation}
where $A_\mu {}^A$ is the Yang-Mills gauge field, $z^\alpha$ are (complex) scalar fields, $\mathcal{K}(z, \overline{z})$ is the K\"ahler potential and $P_A$ is a momentum map or Killing potential, which encodes the non-Abelian gauge transformations on the scalars and may also include Fayet-Iliopoulos terms.

The notation  $\mathcal{L}_0 $, $\mathcal{L}_{1/2} $ and $\mathcal{L}_{1} $
denote respectively the 
kinetic terms of spin $0,\frac{1}{2},1$ fields  in (\ref{Poincact})~\cite{Ferrara:2010yw}:
\begin{eqnarray}
&&  \mathcal{L}_0  =  -\frac1{4\Phi }(\partial _\mu \Phi )(\partial ^\mu \Phi )
  +\frac13g_{\alpha \bar\beta }\Phi (\hat\partial _\mu z^\alpha )\,(\hat\partial ^\mu\bar z^{\bar\beta })
  \,, \nonumber \\
 && \mathcal{L}_{1/2}  = -\ft12\tilde g_{\alpha \bar\beta }
   \bar \chi ^{\bar\beta } \PRa \slashed{D} \chi^\alpha
+ \nonumber \\ 
&& \ft12\Phi  \bar\chi^\alpha\PLa\gamma ^\mu \chi ^{\bar\beta }\hat{\partial }_\mu z^\gamma
\left[-\ft13 g_{\gamma \bar\beta }L_\alpha+\ft14L_{\alpha\gamma }
L _{\bar\beta }-\ft14L_\alpha L _{\gamma\bar\beta }\right]+\hc\,,
  \nonumber \\
 && \mathcal{L}_1 = (\Re f_{AB})\left[ -\ft14 F_{\mu \nu }^AF^{\mu \nu
\,B}-\ft12 \bar\lambda ^A {\slashed{D}}\lambda ^B
\right]\nonumber\\ && +\ft 14\,\rmi\left[(\Im f_{AB})\,
 F_{\mu \nu }^A \tilde F^{\mu \nu \,B}+(\hat{\partial}_\mu\Im f_{AB})\,
\bar\lambda ^A \gamfive\gamma ^\mu \lambda^B\right]\,. \nonumber \\ \label{L1}
\end{eqnarray}
where $f_{AB}(z) $ is a holomorphic kinetic gauge matrix, $ L_\alpha \equiv \partial_\alpha {\rm ln}\left(-\Phi \right)$,
$ L_{\bar \alpha} \equiv \overline{L_\alpha}$, $L_{\alpha \beta} = \partial_{\alpha} L_\beta - \Gamma_{\alpha \beta}^\gamma \, L_\gamma $ 
 and  $g_{\gamma \bar\beta } = -\frac{1}{3} \Phi g_{\alpha \bar \beta} + \frac{1}{4} \Phi \, L_\alpha L_{\bar \beta}$, with $g_{\alpha\bar\beta} = \partial_\alpha \partial_{\bar \beta} \mathcal{K} $ is the K\"ahler metric, with the notation $\partial_\alpha \equiv \frac{\partial}{\partial z^\alpha}$, 
 $\partial_{\bar \alpha} \equiv \frac{\partial}{\partial {\overline z}^{\bar \alpha}}$.

In the notation of \cite{Ferrara:2010yw}, 
the covariant derivatives of the gauginos $\lambda^A$ are defined as:
\begin{eqnarray}
D_\mu \lambda ^A &\equiv &\left( \partial _\mu +\ft14 \omega
_\mu {}^{ab}(e)\gamma _{ab} -\ft32\rmi \mathcal{A}_\mu \gamfive\right)
\lambda ^A-A_\mu ^C\lambda ^B f_{BC}{}^A \,. \nonumber \\
 \label{covderPoinc}
\end{eqnarray}
with $f_{AB}{}^C$  the structure constants of the Non-Abelian gauge group. 

The fermion mass terms, $\mathcal{L}_m $, including gravitino bare mass terms (if any) and the mixed terms $\mathcal{L}_{\mathrm{mix}}$
containing scalars and fermions, including factors of the frame function, are given explicitly in \cite{Ferrara:2010yw}, and again will not be of interest to us in this work. 
We shall be explicitly interested in the 
penultimate of the terms on the right-hand side of Eq.~(\ref{Poincact}), namely: 
\begin{eqnarray}
&& \mathcal{L}_d = \ft18(\Re f_{AB})\bar \psi _\mu \gamma ^{ab}\left( F_{ab}^A+ \widehat F_{ab }^A \right)
\gamma ^\mu \lambda ^B  \nonumber\\
&&+\frac{1}{\sqrt{2}}\left\{\bar \psi_\mu\PLa\gamma ^\nu \gamma^\mu\chi ^\alpha
\left[ (-\ft13\Phi ) g_{\alpha \bar\beta }\hat{\partial }_\nu\bar z^{\bar\beta }+\ft14 L _\alpha
\partial _\nu \Phi
   \right] \right. \nonumber\\
&&\left.\ - \ft1{4}f_{A B\,\alpha } \bar \chi^\alpha \PLa\gamma ^{ab}
\widehat
F_{ab}^{-A } \lambda ^B-\ft13\Phi L_\alpha \bar\chi^\alpha \PLa\gamma ^{\mu \nu} D_\mu \psi_\nu+\hc\right\}\,, \nonumber \\
 \label{parts} 
\end{eqnarray}
where
 \begin{eqnarray}\label{fhat} \widehat{F}_{ab}{}^A\equiv e_a{}^\mu e_b{}^\nu\left( 2\partial _{[\mu }A_{\nu
]}^A +g f_{BC}{}^AA^B_\mu A^C_\nu+\bar \psi _{[\mu } \gamma_{\nu
]}\lambda ^A\right) ~. 
\end{eqnarray}
The explicit
expression for the 4-fermion terms $\mathcal{L}_{4\mathrm{f}}$, which also contain  a significant dependence on the
frame function $\Phi$ and its derivatives, 
will be presented below. Such four-fermion terms, in particular four-gravitino ones, have been argued in 
\cite{Ellis:2011mz} to play an important r\^ole in some variants of the above class of conformal supergravity models, which can characterise certain low-energy limits of superstring theories, in which the frame function $\Phi$ may be identified with the dilaton-axion complex 
superfield, $\frac{\Phi}{3} \equiv \frac{1}{\kappa^2} \, e^{-2\varphi}$. Such models  
can serve as prototypes in which the Deser-Zumino~\cite{Deser:1977uq} mechanism for \emph{dynamical } breaking of local supersymmetry (supergravity) scenario is realised explicitly. The important point to notice in these class of theories is the presence of the frame function $\Phi$ in front of the four-gravitino terms. These  imply that, depending on the value of $\Phi$, assumed to be stabilised appropriately by rolling to the minimum of an appropriate dilaton potential (generated by \emph{e.g}. by string loops, in case one embeds such conformal SUGRA models to string theory, or other ways of breaking the scale symmetry), the effective coupling of the four-gravitino interactions can be much larger than the gravitational coupling. Indeed, in the Einstein frame (denoted by a suffix E), the graviton field (in the original Jordan frame, denoted by a suffix J)  is redefined by means of 
\begin{equation}\label{joreinst}
g_J^{\mu\nu} = e^{-2\varphi} g_E^{\mu\nu} = \left(\frac{\kappa^2 }{3}\, \Phi \right)_{\rm bosonic} g_E^{\mu\nu} ~, \quad  (\Phi |_{\rm bosonic} > 0 ) ~,
\end{equation}
so that the curvature term in the target-space supergravity action has the canonical form, with coefficient the gravitational coupling $\kappa^2 = 8 \pi \, G_N$. with $G_N$ Newton's (four-dimensional) gravitational constant.

In the Einstein frame, where we shall work for the purposes of this work, the gravitational part of the effective conformal supergravity action, including the fermionic torsion induced four-gravitino terms reads
 \begin{eqnarray}\label{confsugra2}
&& \mathcal{L}^E  (e^E)^{-1} =  -\frac{1}{2\kappa^2} R^E(e^E)  + \frac{1}{2} \epsilon^{\mu\nu\rho\sigma} \overline{\psi '}_\mu  \gamma_5 \gamma_\nu D^E_\rho {\psi'}_\sigma -  \nonumber \\ && e^{2\varphi}\,V^E
 +  \frac{11 \kappa^2}{16} e^{-2\varphi} \left[ (\overline{\psi'}_\mu  {\psi'}^\mu )^2  -  (\overline{\psi'}_\mu  \gamma_5 {\psi'}^\mu )^2 \right]
+ \nonumber \\ && \frac{33}{64} \kappa^2 e^{-2\varphi} \, \left(\overline{\psi'}^\rho \gamma_5  \gamma_\mu {\psi'}_\rho  \right)^2  + \dots   =
\nonumber \\
 && -\frac{1}{2\kappa^2} R^E(e^E)  + \frac{1}{2} \epsilon^{\mu\nu\rho\sigma} \overline{\psi '}_\mu  \gamma_5 \gamma_\nu D^E_\rho {\psi'}_\sigma -  
 \nonumber \\ && e^{2\varphi}\,V^E
 +    \rho^2(x)  +  \frac{\sqrt{11}}{2}\kappa \rho (x) e^{-\varphi}\, \left(\overline{\psi'}_\mu {\psi'}^\mu \right)  + 
 \nonumber \\ && \pi^2(x)  +  \frac{\sqrt{11}}{2}e^{-\varphi}\, \kappa \, i \pi (x)  \left(\overline{\psi'}_\mu \gamma_5 {\psi'}^\mu \right) + \nonumber \\ &~& \frac{\sqrt{33}}{2} \kappa \, e^{-\varphi}\, i \lambda^\nu \left(\overline{\psi'}^\rho \gamma_5  \gamma_\nu {\psi'}_\rho \right) + \dots ~,
 \end{eqnarray}
where $R(e)$ denotes the cirvature term with respect to the torsion-free spin connection,  $\psi'_\mu$ denotes the canonically-normalised gravitino with standard kinetic term as in $N=1$ supergravity,
\begin{equation}\label{gravinodef}
\psi ' _\mu =  e^{\varphi} \, \psi_\mu 
\end{equation}
while the $\dots$ denote structures, including auxiliary fields, that are not of direct interest to us here. In writing
(\ref{confsugra2}) we have expanded the four-gravitino terms into detailed structures to exhibit explicitly the terms that
generate masses, and we linearise the four-gravitino terms. The condensate of interest to us is the v.e.v. of the 
linearizing field $\rho (x)$.

The reader should notice that the coefficients of the gravitino-$\rho$ interaction terms in (\ref{confsugra2}) contain dilaton-dependent factors 
$\sim e^{-\varphi}$, and are thus proportional, not to $\kappa$, but to:
\begin{equation}\label{kaptilde}
{\tilde \kappa} \equiv \kappa e^{-\varphi}~. 
\end{equation}
A one-loop analysis shows that the effective potential for the condensate $\rho$ field acquires a minimum at~\cite{Ellis:2011mz}
\begin{equation}
\rho_{\rm min} =   \langle \rho  \rangle   = \pm 0.726 
\end{equation}
at which it vanishes. The gravitino mass term, then, in (\ref{confsugra2}) then takes 
the following form:
\begin{equation}
-m_{3/2} {\overline \psi'}_\mu \Gamma^{\mu\nu} {\psi'}_\nu = 
-\frac{1}{2} m_{3/2} {\overline \psi'}_\mu {\psi'}^\mu ,
\end{equation}
with the dynamically-generated gravitino mass of order 
\begin{equation}\label{gravinomassfinal}
m_{3/2} = \sqrt{11} {\tilde \kappa}^{-1} \rho_{\rm min} = 2.408\, \kappa^{-1} e^\varphi  = \frac{2.408}{\sqrt{8\pi}} \, e^\varphi \, M_P~.
\end{equation}
For large negative values of the v.e.v. $\langle \varphi \rangle < 0$ the resulting gravitino mass is mauch smaller than the Planck scale, and thus the 
effective coupling $\tilde \kappa$ (\ref{kaptilde}) is much larger than the gravitational coupling $G_N$. 

This implies that quantum gravitational corrections to the Minkowski space-time background, on which the above minimisation of the effective potential has been considered are not strong enough to destabilise (at least quickly) the gravitino condensate, unlike the case of standard N=1 supergravity~\cite{Buchbinder:1989gi}. This prompted the authors of \cite{Ellis:2011mz} to consider large positive values of 
\begin{equation}\label{largevalue}
\langle e^{-2\varphi} \rangle \equiv \frac{1}{3} \kappa^2 \langle \Phi |_{\rm bosonic} \rangle  \gg 1~,
\end{equation}
and discuss their relevance to the above-mentioned scenario of dynamical \emph{metastable} breaking of local Sipersymmetry (SUSY) and generation of gravitino mass.

It is the point of this article to examine the constraints implied by such an assumption on the dark sector of the Universe in the case of the NMSSM embedded in this conformal supergravity framework. However, we shall not only restrict ourselves to the case of negative expectation values of the dilaton, but we shall be more general and also consider the dark-matter phenomenology/cosmology of the case of positive dilaton v.e.v. In this latter case, dynamical breaking of local SUSY is not possible in view of the destabilising effects of the graviton fluctuations, nevertheless one assumes 
conventional breaking of SUSY, e.g. through gluino condensation~\cite{nilles}, which is then communicating to the gravity sector to result in a non-trivial gravitino mass term $m_{3/2}$.

\section{Decay processes Involving Gravitinos and Associated Constraints \label{sec:3}} 

We first notice that, in the Einstein frame, one has to first normalise the kinetic terms of the scalars $z^\alpha$ and $\chi$ and $\lambda$ fermions, by appropriate redefinitions involving the 
frame function $\Phi$.  In particular the gauginos should be renormalised in the Einstein frame as
\begin{equation}\label{ginodef}
\lambda ' = e^{2\varphi} \, \lambda  
\end{equation}
in order to acquire canonical kinetic terms. 
On the other hand the gauge terms of the spin-1 part $\mathcal{L}_1$ (\ref{L1}) are already in canonical form, 
in view of the conformal nature of the (Maxwell-like) kinetic terms for the Yang Mills fields. 
Here we consider the case 
$f_{AB}$= const = 1. 

Taking (\ref{gravinodef}) and (\ref{ginodef}) into account, as well as the fact that the Yang-Mills gauge fields are not renormalised in the Einstein frame by 
the frame function $e^{-2\varphi}$, we may write the first term of the right-hand side of (\ref{parts}) in the Einstein frame as:
\begin{equation}\label{lde}
( {L}_d )_E =        \ft18(\Re f_{AB})\bar \psi '_\mu \gamma ^{ab}\left( {F^{\prime}}_{ab}^A+ \widehat{F^\prime}_{ab }^A \right)
\gamma ^\mu \lambda ^{\prime \,B } ~.
\end{equation}
This term is responsible for the gravitino--gaugino--gauge-boson interaction and it does not transformed going to the new frame.

Tying to recap the transformations we have 
 \begin{eqnarray}
\psi ' _\mu &=&  e^{\varphi} \, \psi_\mu   \nonumber \\
\lambda ' &= &e^{2\varphi} \, \lambda  \, .
\label{eq:trans}
 \end{eqnarray}
The scalar and vector fields don't change. Using these we can 
calculate how the interactions that are relevant to the gravitino decays change due to the dilation presence.
In particular we are interested in  the $\psi \, \chi \, Z (\gamma)$ decays, where with $\chi$ we denote 
the neutralino that is the LSP in our model.
Doing so, we may then  consider the terms $\mathcal{L}_d$ (\ref{parts}) in order to compute the decay rate of the massive gravitino field ${\tilde G}$  into , say, a neutralino $\chi$ in the NMSSM and a $Z$ gauge boson:
$$ {\tilde G} \quad \rightarrow \chi + Z ~. $$
The so affected gravitino decay rate, will in turn affect the Dark Matter relic density (assumed to be dominated by neutralinos in NMSSM) and
this may imply stronger Big Bang Nucleosynthesis (BBN) constraints.
It is therefore important that detailed cosmological studies of such dilaton
extended mSUGRA models are performed.

In the usual mSUGRA gravitino satisfies the Rarita-Schwinger (RS)  eqs.
 \begin{eqnarray}
\gamma^\mu \psi_\mu (x)&=&0  \nonumber \\
(i\slashed{\partial}-m) \psi_\mu (x)&=&0
\label{eq:RS}
 \end{eqnarray}
 which result from the RS action~\cite{Rarita}
 \begin{equation}
 \mathcal{L}=-\frac{1}{2}\epsilon^{\mu \nu \rho\sigma} \bar{\psi}_\mu \gamma_5 \gamma_\nu \partial_\rho \psi_\sigma 
                         -\frac{1}{4} m_{3/2} \bar{\psi}_\mu [\gamma^\mu,\gamma^\nu] \psi_\nu \,.
 \end{equation} 
The RS action including the dilaton effects can be written as
 \begin{equation}\label{gvinomassterm}
 \mathcal{L}' =-\frac{1}{2}\epsilon^{\mu \nu \rho\sigma} \bar{\psi}'_\mu \gamma_5 \gamma_\nu \partial_\rho \psi'_\sigma 
                         -\frac{1}{4} m'_{3/2}  \bar{\psi}'_\mu [\gamma^\mu,\gamma^\nu] \psi'_\nu
 \end{equation}
where $\psi'_\mu \to e^ \varphi   \psi_\mu$, and the relation between $m'_{3/2}$ to  $m_{3/2}$
is 
\begin{equation} \label{gvinomassterm2}
m_{3/2}^\prime =  e^\varphi\, m_{3/2}
\end{equation}
so for $\langle \varphi \rangle < 0 $, which is the physical case in several of the backgrounds discussed  in order to allow for dynamical breaking of local 
SUSY, the gravitino mass in the conformal supergravity scenario will be smaller than the corresponding one in the normal SUGRA. 

 In the NMSSM~\cite{NMSSM} the neutralino field can be written as:
 \begin{equation}
\chi = N_{11} \tilde{B}+ N_{12} \tilde{W}^3_0+N_{13} \tilde{H}^1_0 + N_{14} \tilde{H}^2_0 + N_{15} \tilde{S} \, ,
 \label{chinmssm}
\end{equation}
where $N_{ij}$ are the elements of the $5 \times 5$ neutralino diagonalizing matrix.
 If $N_{11}$ dominates the sum  $ \sum_{i=1,5}N_{i1}^2=1 $, 
  then the lightest neutralino is bino-like. On the other hand,  if $N_{15}$ is dominant then the neutrslino is 
singlino-like. The latest data of the LHC experiments, indicating a Higgs boson mass in the ballpark of 125 GeV~\cite{LHC}, 
combined with other experimental data from B-physics and direct dark matter searches, seem to 
disfavor the singlino case~\cite{Kowalska:2012gs}. Thus, in the following it will be assumed that the lightest neutralino 
is mainly bino.   In this case, the dominant two body decay channels for gravitino are ${\tilde G} \to \gamma \chi$ and ${\tilde G} \to Z \chi$.
Nevertheless, even in the singlino-like case those channels, especially the $\gamma \chi$,  are 
 dominant,  mainly due to the large  available phase space. Therefore,  our assumption that the light neutralino is mainly bino is sufficiently generic.

We note in passing at this point that, as shown in~\cite{dilatondm}, 
 a time-dependent (cosmological) dilaton (which can run with the cosmic time before BBN) can reduce considerably the  neutralino relic density, thereby 
 increasing the cosmologically-allowed available parameter space of SUSY even beyond the LHC reach. 
In this article we ignore such effects, focusing on the gravitino interactions exclusively, and assuming that the dilaton in our case has been stabilised to its vacuum expectation value at the scale of SUSY breaking  (or at least it is approximately constant during a cosmological epoch). However, even for stabilised dilatons, in the Einstein frame, the neutralino-pair annihilation processes are affected by the conformal couplings, in particular are enhanced for large $e^\varphi > 1$. Such enhancement may reduce the relic abundances already in the constant dilaton case. 
We  postpone  a comprehensive study of such effects  on  
 gravitino decays and the neutralino  dark matter abundance for a future study.  
 
Below we shall consider two cases: one,  in which the neutralino is the stable Dark Matter candidate and the gravitino is heavier, thus unstable, and the other, in which the gravitino is cosmologically stable and thus constitutes the dark matter candidate. We commence our discussion from the former case. 

In such a case, the  formula for the decay width ${\tilde G} \to \gamma \chi$ without the dilaton effects, reads as~\cite{grdm}:
\be
\Gamma_{\gamma \chi} = \frac{1}{16\, \pi} \frac{\overline{|\mathcal{M_\gamma}|^2}}{\mgr} \mathcal{F}(\mgr,m_\chi,0),
\label{eq:gamg}
\ee
where the spin average amplitude squared is 
\be
\overline{|\mathcal{M_\gamma}|^2} = \frac{B_\gamma^2}{6 M_P^2} \frac{1}{\mgr^2}\, ( m_{3/2}^2-m_\chi^2)^2   ( 3 \, m_{3/2}^2+m_\chi^2 )
\ee
and the kinematical factor is defined as
\bea
\mathcal{F}(m_0,m_1,m_2) &=& \frac{1}{m_0^2} \left[  \left( m_0^2-(m_1+m_2)^2 \right) \right.    \nonumber \\ 
      &\times&      \left.    \left( m_0^2-(m_1-m_2)^2 \right) \right]^{1/2} \,.
\eea 
On the other hand, taking into account the dilaton effects, \emph{i.e.} considering the corresponding process in the conformal SUGRA model in the Einstein frame, the  corresponding width becomes
\be
 \Gamma'_{\gamma \chi} =  \frac{1}{16\, \pi} \frac{\overline{|\mathcal{M_\gamma'}|^2}}{c\,\mgr} \mathcal{F}(c\,\mgr,m_\chi,0)\, ,
      \label{eq:gamgp}
\ee
where
\bea
&& \overline{|\mathcal{M_\gamma'}|^2} =  \frac{B_\gamma^2}{12\, M_P^2} \, \frac{1}  { \mgr^2} \,  c^2\, ( m_{3/2}^2-m_\chi^2)^2   \nonumber \\
     &&\times  \left(6 \, c^4 \, m_{3/2}^2+\left(c^2+1\right) m_\chi^2\right) \,. 
\eea

Above  it was defined  $c =  e^\varphi$.  Notice that putting  $c=1$ we recover the result of Eq.~(\ref{eq:gamg}).
The factor $B_\gamma$ is related to the bino ($\tilde{B}$) and neutral wino ($\tilde{W}^3_0$) components of the neutralino, that is
$B_\gamma =N_{i1} \cos \theta_W+ N_{i2}\sin\theta_W $, where  $\theta_W$ is
the electroweak mixing angle.

For the channel ${\tilde G} \to Z \chi$ in the standard (dilaton-free) SUGRA the width reads as:
\be
\Gamma_{Z \chi} = \frac{1}{16\, \pi} \frac{\overline{|\mathcal{M}_Z |^2}}{\mgr} \mathcal{F}(\mgr,m_\chi,M_Z),
\label{eq:gamz}
\ee
where 
\bea
 \overline{ |\mathcal{M}_Z |^2} &=  &\frac{B_Z^2}{6 M_P^2} \frac{1}{\mgr^2}\, 
 \left[ 3 \mgr^6-\mgr^4 \left(5m_\chi^2 + M_Z^2\right)  \right. \nonumber \\
 & + &  12 \mgr^3m_\chi M_Z^2 + \mgr^2 \left(m_\chi^4-M_Z^4\right)   \nonumber \\
 &+&  \left.   \left(m_\chi^2-M_Z^2\right)^3\right] \, ,
\eea
where $B_Z = - N_{i1} \sin \theta_W + N_{i2} \cos \theta_W $.
In the case of dilaton the same width becomes 
\be
\Gamma'_{Z \chi} = \frac{1}{16\, \pi} \frac{\overline{|\mathcal{M'}_Z |^2}}{c\,\mgr} \mathcal{F}(c\, \mgr,m_\chi,M_Z) \, ,
 \label{eq:gamzp}
\ee
where
\bea
&& \overline{|\mathcal{M'}_Z|^2} =  \frac{B_Z^2}{24\, M_P^2} \, \frac{1}  { \mgr^2} \,   \nonumber \\
 &\times&  c^2 \left[-2 \left(c^2+1\right) M_Z^6+M_Z^4 \left(6 \left(c^2+1\right) m_\chi^2 \right. \right.  \nonumber \\
 &+&   \left. \left(-6 c^4+c^2+1\right) \mgr^2\right)  \nonumber \\
&+& 2   (\mgr-m_\chi)^2 (\mgr+m_\chi)^2 \left(6 c^4 \mgr^2+\left(c^2+1\right) m_\chi^2\right)   \nonumber \\
&+& M_Z^2 \left(-6 \left(c^2+1\right) m_\chi^4 \right.  \nonumber \\
&+ & \left(-6 c^4+c^2+1\right) \mgr^4+6 c \left(12 c^4-3 c^2-1\right) \mgr^3 m_\chi   \nonumber \\
&+& 3 \left. \left.  \left(-2 c^4+c^2+1\right) \mgr^2 m_\chi^2\right) \right] \, .
\eea
It is worth noticing that  $\Gamma_{Z \chi}$  goes to $\Gamma_{ \gamma \chi}$ in the limit $M_Z \to 0$, and the same holds
for $\Gamma'$.

%%%%%%%%%%%
\begin{figure}[th]
\begin{center}
\includegraphics[width=\linewidth]{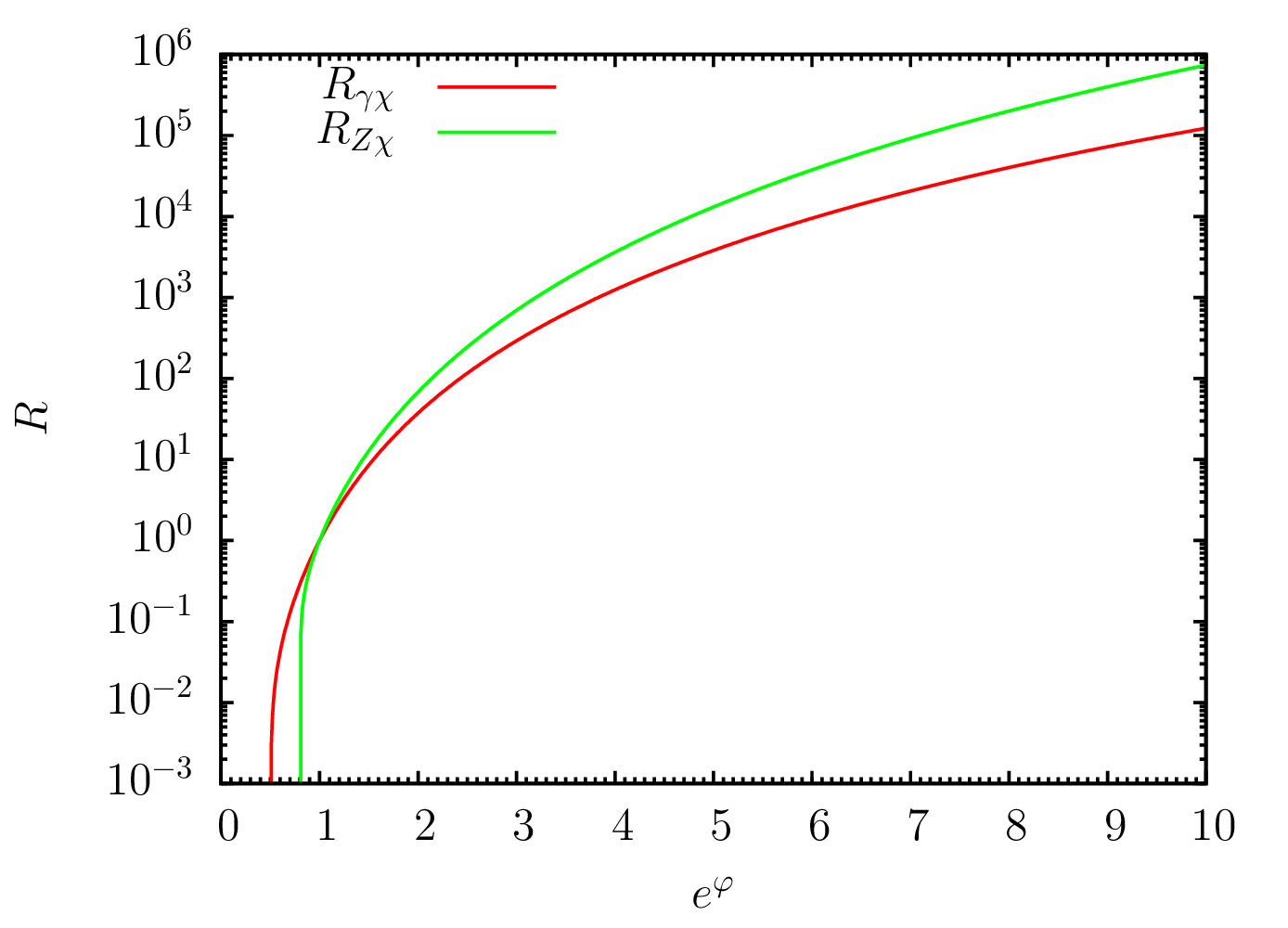}
\end{center}
\caption[]{
The ratio $R=\Gamma' /\Gamma $ for the  channels  ${\tilde G} \to  \chi \, \gamma $ and ${\tilde G} \to  \chi \, Z$,  as a function of $e^\varphi$, for the 
neutralino dark matter case.
}
\label{fig:ndm}  
\end{figure}
%%%%%%%%%%%%%%%%%%%%%%%%%%

We start our numerical analysis discussing models where dark matter consists of neutralinos and the gravitino 
is unstable. In this case decays of the gravitinos to neutralinos can be an important constraint, affecting significantly  the BBN predictions.
In Fig.~\ref{fig:ndm} we present the ratio $R=\Gamma' /\Gamma $ for the processes ${\tilde G} \to \gamma \, \chi$ and ${\tilde G} \to Z\, \chi $ as a function of $e^\varphi$. To make this figure we use the numerical 
values $m_{3/2}=300$ GeV, $m_\chi=150$ GeV, and for these  the  two body decay widths for the 
dominant channels (involving $\gamma$) are $1.4 \times 10^{-32} $ GeV and  $3.3 \times 10^{-33} $ GeV for the  $\gamma \, \chi$ and $Z \, \chi$, respectively.
This yields a gravitino lifetime  $\sim \, 4.7\times 10^{7}$ s.

We first concentrate in the region of the  Fig.~\ref{fig:ndm} where $c > 1$,
 for which although dynamical generation of gravitino mass and thus breaking of local SUSY may \emph{not} occur~\cite{Ellis:2011mz,Buchbinder:1989gi} nevertheless, as mentioned above, one may assume a more or less conventional mechanism~\cite{nilles} for SUSY 
 breaking and the generation of a gravitino mass term $m_{3/2}$, (\ref{gvinomassterm}), 
(\ref{gvinomassterm2}).  In this case, there is an enhancement of the ratio $R=\Gamma' /\Gamma $, where the prime denotes the width for the conformal SUGRA case, where dilaton effects are taken into account. 
An important observation concerns the fact  that for $c < 1 $, which is the case where dynamical (metastable) breaking of local SUSY mass be possible, according to the arguments of \cite{Ellis:2011mz} reviewed above (\emph{cf}.
 discussion leading to Eq.~(\ref{gravinomassfinal})) the decay of gravitino to photons and neutralinos is kinematically \emph{forbidden}. 
 This case will face important constraints from BBN which will be discussed below. 

%%%%%%%%%%%
\begin{figure}[th]
\begin{center}
\includegraphics[width=\linewidth]{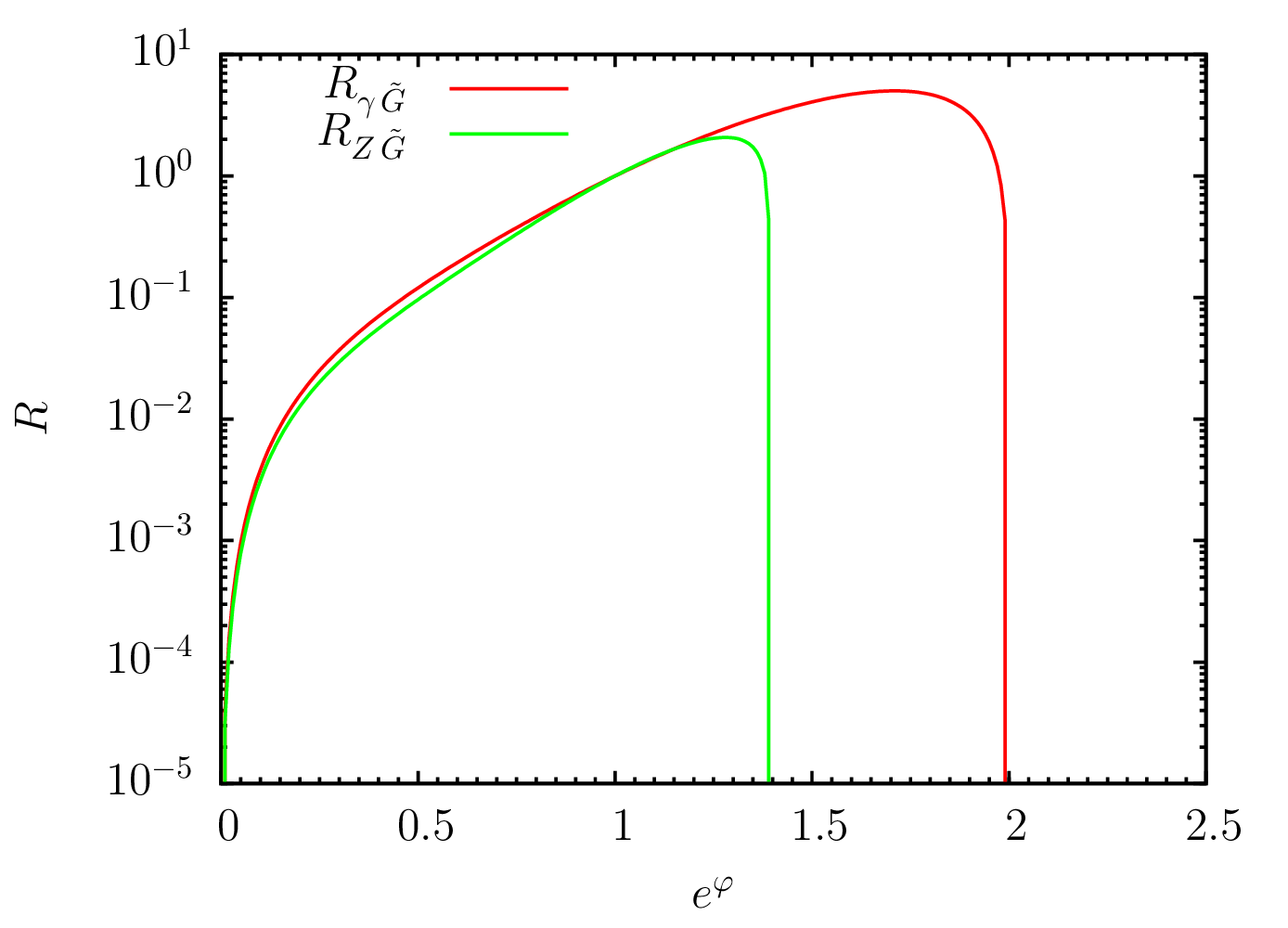}
\end{center}
\caption[]{
The ratio $R=\Gamma' /\Gamma $ for the  channels  $  \chi \to {\tilde G}  \, \gamma$ and $\chi \to    {\tilde G} \, Z $,  as a function of $e^\varphi$
for the gravitino dark matter case.
}
\label{fig:gdm}  
\end{figure}
%%%%%%%%%%%%%%%%%%%%%%%%%%

 Concerning the BBN constraints, we know that they become important for lifetime of the unstable particle  $ \tau \gtrsim 10^2$ s~\cite{bbn}. 
 Since  $ {\tau}' =\tau / R $ and the gravitino lifetime without the dilaton effects is $\mathcal{O}(10^7 )$ s,
 one observes that with $R \gtrsim 10^5$ one avoids all the important BBN constraints practically for any gravitino mass. 
 Using Fig.~\ref{fig:ndm} we understand that this happens for $ e^\varphi  \gtrsim 9$ or 10. 
 For smaller values of $ e^\varphi $ the BBN appears to become important, since
 the lifetime of the gravitino approaches its original value $10^7$ s as $e^\varphi \to 1$, but  its abundance is still 
 enhanced by the factor $e^\varphi$. On the other hand, values for $e^\varphi < 1$ enhance dramatically the 
 gravitino lifetime, leading to the \emph{exclusion}  of this range of the scale factor, being incompatible with the BBN constraints.
We have checked numerically different values of gravitino  mass  up to TeV, but our results  
remain basically the same. That is,  one finds that for $ e^\varphi $ in the range above 8 or 10,  the BBN constraints can be avoided.

 In addition,  we discuss also the complementary case where the gravitino is the LSP. In this case one studies the reverse 
 processes $  \chi \to {\tilde G}  \, \gamma$ and $\chi \to    {\tilde G} \, Z $, from the point of view of the constraints induced by BBN.
 To compute the corresponding decay widths we use the fact that the amplitudes squared $\overline{| \mathcal{M}_{\gamma,Z} |^2}$,
 both for the standard and the dilaton cases, are the same as before, due to the assumed CPT invariance and unitarity. On the other hand,  one has to interchange $m_\chi$ and $\mgr$ in 
 $\mathcal{F}$ and in the denominators of Eqs.~(\ref{eq:gamg}), (\ref{eq:gamgp}), (\ref{eq:gamz}), and ~(\ref{eq:gamzp}). 
 Doing so, we plot in Fig.~\ref{fig:gdm} the ratio $R=\Gamma' /\Gamma $ for these reverse processes. 
   The numerical values we use are $m_{3/2}=150$ GeV, $m_\chi=300$ GeV, exactly the 
  reverse case of Fig.~\ref{fig:ndm}. With this choice, the  neutralino  lifetime is now  $\mathcal{O} (10^{7})$ s.
We thus  see that the kinematically allowed region happens  for values of $c=e^{\varphi} < 2$. This region 
depends on the choice neutralino and gravitino masses, but the attained values of $R$ are not sensitive 
to this. One observes that in this case $R \lesssim 10$.  Thus in this case the dilaton effects can not be used 
to relax the BBN constraints.

\section{Conclusions and Outlook \label{sec:4}}

In this article we have discussed the effects of stabilised dilatons on processes involving unstable particles, including 
gravitinos, that may affect BBN  in conformal SUGRA models, incorporating the NMSSM in their spectra. 
We have found that, in the case where the gravitino is unstable, and the neutralino plays the role of dark matter, there are regions of the scale (dilaton) factor $e^\varphi > 8 $ in which the BBN constraints can be avoided altogether. Moreover, in such regimes the neutralino dark matter abundances may be diluted thereby avoiding the cosmological and particle physics  constraints on SUSY matter at current colliders, including LHC. Unfortunately, this case seems not to favour dynamical SUGRA breaking (due to quantum-gravity instabilities), and therefore one has to assume more-or-less conventional breaking of SUSY and its communication to the gravitational sector. 
On the other hand, if the gravitino is cosmologically stable, playing the role of dark matter, the BBN constraints are very restrictive in the full (kinematically-allowed) range of the scale (dilaton) factor $e^{\varphi}$.   

We have not discussed in detail time-dependent dilaton effects that involve running scale factors up to BBN, which are known to reduce significantly the dark matter abundances.
Such effects, when combined with the dilaton effects on the gravitino decays considered here, may change significantly the cosmology and phenomenology of such conformal SUGRA models. We plan to return to these interesting issues in a forthcoming publication. 

\section*{Acknowledgements}

The work of N.E.M. is supported in part by STFC (UK) under the research grant ST/J002798/1 and by the London Centre for
Terauniverse Studies (LCTS), using funding from the European Research
Council via the Advanced Investigator Grant 267352; this latter grant also supported visits 
by  V.C.S. to the CERN TH Division, which he thanks for its hospitality and where this work was initiated. The work of V.C.S. is supported
by Marie Curie International Reintegration grant SUSYDM-PHEN, MIRG-CT-2007-203189.


\begin{thebibliography}{99}

%\cite{Ferrara:2010yw}
\bibitem{Ferrara:2010yw} 
  S.~Ferrara, R.~Kallosh, A.~Linde, A.~Marrani and A.~Van Proeyen,
  %``Jordan Frame Supergravity and Inflation in NMSSM,''
  Phys.\ Rev.\ D {\bf 82}, 045003 (2010)
  [arXiv:1004.0712 [hep-th]].
  %%CITATION = ARXIV:1004.0712;%%

%\cite{Ferrara:2010in}
\bibitem{Ferrara:2010in} 
  S.~Ferrara, R.~Kallosh, A.~Linde, A.~Marrani and A.~Van Proeyen,
  %``Superconformal Symmetry, NMSSM, and Inflation,''
  Phys.\ Rev.\ D {\bf 83}, 025008 (2011)
  [arXiv:1008.2942 [hep-th]].
  %%CITATION = ARXIV:1008.2942;%%

%\cite{Bezrukov:2007ep}
\bibitem{Bezrukov:2007ep} 
  F.~L.~Bezrukov and M.~Shaposhnikov,
  %``The Standard Model Higgs boson as the inflaton,''
  Phys.\ Lett.\ B {\bf 659}, 703 (2008)
  [arXiv:0710.3755 [hep-th]].
  %%CITATION = ARXIV:0710.3755;%%

%\cite{Bezrukov:2009db}
\bibitem{Bezrukov:2009db} 
  F.~Bezrukov and M.~Shaposhnikov,
  %``Standard Model Higgs boson mass from inflation: Two loop analysis,''
  JHEP {\bf 0907}, 089 (2009)
  [arXiv:0904.1537 [hep-ph]].
  %%CITATION = ARXIV:0904.1537;%%




%\cite{Bezrukov:2010jz}
\bibitem{Bezrukov:2010jz} 
  F.~Bezrukov, A.~Magnin, M.~Shaposhnikov and S.~Sibiryakov,
  %``Higgs inflation: consistency and generalisations,''
  JHEP {\bf 1101}, 016 (2011)
  [arXiv:1008.5157 [hep-ph]].
  %%CITATION = ARXIV:1008.5157;%%

 %\cite{GarciaBellido:2011de}
\bibitem{GarciaBellido:2011de} 
  J.~Garcia-Bellido, J.~Rubio, M.~Shaposhnikov and D.~Zenhausern,
  %``Higgs-Dilaton Cosmology: From the Early to the Late Universe,''
  arXiv:1107.2163 [hep-ph].
  %%CITATION = ARXIV:1107.2163;%%


%\cite{Einhorn:2009bh}
\bibitem{Einhorn:2009bh} 
  M.~B.~Einhorn and D.~R.~T.~Jones,
  %``Inflation with Non-minimal Gravitational Couplings in Supergravity,''
  JHEP {\bf 1003}, 026 (2010)
  [arXiv:0912.2718 [hep-ph]].
  %%CITATION = ARXIV:0912.2718;%%

 
%\cite{AlvarezGaume:2010rt}
\bibitem{AlvarezGaume:2010rt} 
  L.~Alvarez-Gaume, C.~Gomez and R.~Jimenez,
  %``Minimal Inflation,''
  Phys.\ Lett.\ B {\bf 690}, 68 (2010)
  [arXiv:1001.0010 [hep-th]].
  %%CITATION = ARXIV:1001.0010;%%

%\cite{AlvarezGaume:2011xv}
\bibitem{AlvarezGaume:2011xv} 
  L.~Alvarez-Gaume, C.~Gomez and R.~Jimenez,
  %``A Minimal Inflation Scenario,''
  JCAP {\bf 1103}, 027 (2011)
  [arXiv:1101.4948 [hep-th]].
  %%CITATION = ARXIV:1101.4948;%%
  

\bibitem{NMSSM} C.~Panagiotakopoulos and K.~Tamvakis,
  %``Stabilized NMSSM without domain walls,''
  Phys.\ Lett.\ B {\bf 446}, 224 (1999)
  [hep-ph/9809475];
  %%CITATION = HEP-PH/9809475;%%
  M.~Maniatis,
  %``The Next-to-Minimal Supersymmetric extension of the Standard Model reviewed,''
  Int.\ J.\ Mod.\ Phys.\ A {\bf 25}, 3505 (2010)
  [arXiv:0906.0777 [hep-ph]];
  %%CITATION = ARXIV:0906.0777;%%  
For a  review see:  U.~Ellwanger, C.~Hugonie and A.~M.~Teixeira,
  %``The Next-to-Minimal Supersymmetric Standard Model,''
  Phys.\ Rept.\  {\bf 496}, 1 (2010)
  [arXiv:0910.1785 [hep-ph]] and references therein.
  %%CITATION = ARXIV:0910.1785;%%
  



%\cite{Ellis:2011mz}
\bibitem{Ellis:2011mz} 
  J.~Ellis and N.~E.~Mavromatos,
  %``On the Role of Space-Time Foam in Breaking Supersymmetry via the Barbero-Immirzi Parameter,''
  Phys.\ Rev.\ D {\bf 84}, 085016 (2011)
  [arXiv:1108.0877 [hep-th]].
  %%CITATION = ARXIV:1108.0877;%%

%\cite{Deser:1977uq}
\bibitem{Deser:1977uq} 
  S.~Deser and B.~Zumino,
  %``Broken Supersymmetry and Supergravity,''
  Phys.\ Rev.\ Lett.\  {\bf 38}, 1433 (1977).
  %%CITATION = PRLTA,38,1433;%%

%\cite{Buchbinder:1989gi}
\bibitem{Buchbinder:1989gi}
  I.~L.~Buchbinder and S.~D.~Odintsov,
  %``IS DYNAMICAL SUPERSYMMETRY BREAKING IN N=1 SUPERGRAVITY POSSIBLE?,''
  Class.\ Quant.\ Grav.\  {\bf 6}, 1955 (1989).
  %%CITATION = CQGRD,6,1955;%%

%\cite{Nilles:1983ge}
\bibitem{nilles} see, for instance:
  H.~P.~Nilles,
  %``Supersymmetry, Supergravity and Particle Physics,''
  Phys.\ Rept.\  {\bf 110}, 1 (1984) and references therein.
  %%CITATION = PRPLC,110,1;%%

 
\bibitem{Rarita}
% On a Theory of Particles with Half-Integral Spin
W. Rarita and J. Schwinger, Phys. Rev. \textbf{60}, 61  (1941). 


\bibitem{LHC} G.~Aad {\it et al.}  [ATLAS Collaboration],
  %``Observation of a new particle in the search for the Standard Model Higgs boson with the ATLAS detector at the LHC,''
  Phys.\ Lett.\ B {\bf 716}, 1 (2012)
  [arXiv:1207.7214 [hep-ex]];
  %%CITATION = ARXIV:1207.7214;%%
S.~Chatrchyan {\it et al.}  [CMS Collaboration],
  %``Observation of a new boson at a mass of 125 GeV with the CMS experiment at the LHC,''
  Phys.\ Lett.\ B {\bf 716}, 30 (2012)
  [arXiv:1207.7235 [hep-ex]].
  %%CITATION = ARXIV:1207.7235;%%

\bibitem{Kowalska:2012gs}
  K.~Kowalska, S.~Munir, L.~Roszkowski, E.~M.~Sessolo, S.~Trojanowski and Y.~-L.~S.~Tsai,
  %``The Constrained NMSSM with a 125 GeV Higgs boson -- A global analysis,''
  arXiv:1211.1693 [hep-ph].
  %%CITATION = ARXIV:1211.1693;%%

  
\bibitem{dilatondm}
 A.~B.~Lahanas, N.~E.~Mavromatos and D.~V.~Nanopoulos,
  %``Dilaton and off-shell (non-critical string) effects in Boltzmann equation for species abundances,''
  PMC Phys.\ A {\bf 1}, 2 (2007) 
  [hep-ph/0608153];
  %%CITATION = HEP-PH/0608153;%%
  %``Smoothly evolving supercritical-string dark energy relaxes supersymmetric-dark-matter constraints,''
  Phys.\ Lett.\ B {\bf 649}, 83 (2007) 
  [hep-ph/0612152];
  %%CITATION = HEP-PH/0612152;%%
 A.~B.~Lahanas,
  %``Dilaton dominance in the early Universe dilutes Dark Matter relic abundances,''
  Phys.\ Rev.\ D {\bf 83},  103523
   (2011)   [arXiv:1102.4277 [hep-ph]];
  A.~B.~Lahanas and V.~C.~Spanos,
  %``Dilaton dominance relaxes LHC and cosmological constraints in supersymmetric models,''
  JHEP {\bf 1206}, 089 (2012) 
  [arXiv:1201.2601 [hep-ph]].
  %%CITATION = ARXIV:1102.4277;%%
  
  
  \bibitem{grdm}
  J.~R.~Ellis, K.~A.~Olive, Y.~Santoso and V.~C.~Spanos,
  %``Gravitino dark matter in the CMSSM,''
  Phys.\ Lett.\ B {\bf 588}, 7 (2004) 
  [hep-ph/0312262].
  %%CITATION = HEP-PH/0312262;%%

 \bibitem{bbn}
 M.~Kawasaki, K.~Kohri and T.~Moroi,
  %``Big-Bang nucleosynthesis and hadronic decay of long-lived massive particles,''
  Phys.\ Rev.\ D {\bf 71},  083502 (2005) 
  [astro-ph/0408426];
  %%CITATION = ASTRO-PH/0408426;%%
 R.~H.~Cyburt, J.~Ellis, B.~D.~Fields, F.~Luo, K.~A.~Olive and V.~C.~Spanos,
  %``Nucleosynthesis Constraints on a Massive Gravitino in Neutralino Dark Matter Scenarios,''
  JCAP {\bf 0910}, 021 (2009) 
  [arXiv:0907.5003 [astro-ph.CO]].
  %%CITATION = ARXIV:0907.5003;%%



\end{thebibliography}
\end{document}